\documentclass[12pt,letterpaper,oneside]{article}

\usepackage{cite}
\usepackage{bm}
\usepackage[cmex10]{amsmath}
\usepackage{array}
\usepackage{fixltx2e}
\usepackage{amsfonts}
\usepackage{color}
\usepackage{amsmath}
\usepackage{amsthm}
\usepackage{amssymb}
\usepackage{t1enc}
\usepackage{alltt}
\usepackage{epsfig}
\usepackage{mathrsfs}
\usepackage{graphicx,ifthen}
\usepackage{multirow}
\usepackage[scaled]{helvet}
\usepackage{booktabs}
\usepackage{wasysym}
\usepackage{arydshln}
\usepackage{enumerate}

\newcommand{\tr}{{\mathrm{tr}}}

\newcommand{\dd}{\,\mathrm{d}}

\makeatletter

\newcommand{\Rmnum}[1]{\expandafter\@slowromancap\romannumeral #1@}

\begin{document}
\title{A note on Bianchi-Don\`{a}'s proof to the variance formula of von Neumann entropy}
\author{Lu Wei\footnote{Department of Electrical and Computer Engineering, University of Michigan - Dearborn, Michigan 48128, USA. Email: luwe@umich.edu}}
\date{\today}

\maketitle
\begin{abstract}
Bianchi and Don\`{a}~\cite{Bianchi19} have recently reported a proof to the variance formula of von Neumann entropy, which was conjectured in~\cite{VPO16} and firstly proved in~\cite{Wei17}. The purpose of this short note is to show that, despite having a different starting point, the subsequent calculations (omitted in~\cite{Bianchi19}) leading to the result are essentially the same as in~\cite{Wei17}.
\end{abstract}

\section{Introduction}
We concisely outline the mathematical formulation and notations of the considered problem, where the physical background can be found in~\cite{Bianchi19,VPO16}. Consider a bipartite quantum system that consists of two subsystems $A$ and $B$ of Hilbert space dimensions $m$ and $n$. For such a system, the eigenvalue density of the reduced density matrix is
\begin{eqnarray}\label{eq:fte}
f\left(\bm{\lambda}\right)\propto~\delta\left(1-\sum_{i=1}^{m}\lambda_{i}\right)\prod_{1\leq i<j\leq m}\left(\lambda_{i}-\lambda_{j}\right)^{2}\prod_{i=1}^{m}\lambda_{i}^{n-m},
\end{eqnarray}
where $0<\lambda_{i}<1$ and $\sum_{i=1}^{m}\lambda_{i}=1$. The entanglement of the bipartite system can be understood from the moments
\begin{equation}\label{eq:m}
\mathbb{E}_{f}\!\left[S^{k}\right],~~~~~~k=1,2,\dots,
\end{equation}
of von Neumann entropy
\begin{equation}\label{eq:vN}
S=-\sum_{i=1}^{m}\lambda_{i}\ln\lambda_{i},~~~~~~S\in\left[0, \ln{m}\right],
\end{equation}
over the environment~(\ref{eq:fte}). It is known that the mean value of von Neumann entropy is given by~\cite{Page93}
\begin{equation}\label{eq:vNm}
\mathbb{E}_{f}\!\left[S\right]=\psi_{0}(mn+1)-\psi_{0}(n)-\frac{m+1}{2n},
\end{equation}
where $\psi_{0}(x)=\dd\ln\Gamma(x)/\dd x$ is the digamma function.

This note concerns the variance of von Neumann entropy, a formula of which was conjectured by Vivo, Pato, and Oshanin~\cite{VPO16} as
\begin{equation}\label{eq:V}
\mathbb{V}\!_{f}\!\left[S\right]=-\psi_{1}\left(mn+1\right)+\frac{m+n}{mn+1}\psi_{1}\left(n\right)-\frac{(m+1)(m+2n+1)}{4n^{2}(mn+1)},
\end{equation}
where $\psi_{1}(x)=\dd^{2}\ln\Gamma(x)/\dd x^{2}$ is the trigamma function. The above formula was firstly proved in~\cite{Wei17} by computing directly the second moment $\mathbb{E}_{f}\!\left[S^{2}\right]$. Recently, Bianchi and Don\`{a} also reported a proof~\cite{Bianchi19} to the variance formula~(\ref{eq:V}), where the starting point is their newly discovered representation~\cite[Eq.~(S29)]{Bianchi19}
\begin{eqnarray}\label{eq:sp}
\mathbb{E}_{f}\!\left[\left(\sum_{i=1}^{m}\lambda_{i}^{r_{1}}\right)\left(\sum_{i=1}^{m}\lambda_{i}^{r_{2}}\right)\right]=\frac{\Gamma(mn)}{\Gamma(mn+r_{1}+r_{2})}\big(\tr\mathbf{X}(r_{1}+r_{2})+\nonumber\\
\tr\mathbf{X}(r_{1})\tr\mathbf{X}(r_{2})-\tr\left(\mathbf{X}(r_{1})\mathbf{X}(r_{2})\right)\big)
\end{eqnarray}
with the entry $x_{ij}$ ($i,j=0,\dots,m-1$) of the $m\times m$ matrix $\mathbf{X}(r)$ being
\begin{equation}\label{eq:xij}
\frac{\Gamma^{2}(r+1)\Gamma(j+1)}{\Gamma(n-m+i+1)}\sum_{k=0}^{m-1}\frac{\left(k!\Gamma(i-k+1)\Gamma(j-k+1)\right)^{-1}\Gamma(n-m+r+k+1)}{\Gamma(r+k-i+1)\Gamma(r+k-j+1)}.
\end{equation}
Clearly, the second moment will be recovered by taking the limits of the derivatives of the representation~(\ref{eq:sp}), i.e.,
\begin{equation}\label{eq:T2vN}
\mathbb{E}_{f}\!\left[S^{2}\right]=\lim_{\substack{r_{1}\to1\\r_{2}\to1}}\frac{\partial^{2}}{\partial r_{2}\partial r_{1}}\mathbb{E}_{f}\!\left[\left(\sum_{i=1}^{m}\lambda_{i}^{r_{1}}\right)\left(\sum_{i=1}^{m}\lambda_{i}^{r_{2}}\right)\right].
\end{equation}
The variance formula~(\ref{eq:V}) was then immediately declared in~\cite{Bianchi19}, where the authors described the calculation procedure of~(\ref{eq:T2vN}) as ``We take the derivatives and the limits with the help of Wolfram's Mathematica''.

The aim of this note is to reveal the details of the calculations in~(\ref{eq:T2vN}). We find that the computation of~(\ref{eq:T2vN}) necessarily involves, albeit in a different order, the bulk of calculations and simplifications as reported in~\cite{Wei17}. Moreover, the computation relies on new summation identities (derived in~\cite{Wei17}) that in fact have not been implemented in Mathematica~\cite{KG}.

\section{Calculations of~(\ref{eq:T2vN})}
Inserting~(\ref{eq:sp}) into~(\ref{eq:T2vN}), the computation can be divided into three parts
\begin{equation}
\mathbb{E}_{f}\!\left[S^{2}\right]=T_{a}+T_{b}+T_{c},
\end{equation}
where
\begin{eqnarray}
T_{a} &=& \Gamma(mn)\lim_{\substack{r_{1}\to1\\r_{2}\to1}}\frac{\partial^{2}}{\partial r_{2}\partial r_{1}}\frac{\tr\mathbf{X}(r_{1}+r_{2})}{\Gamma(mn+r_{1}+r_{2})}, \\
T_{b} &=& \Gamma(mn)\lim_{\substack{r_{1}\to1\\r_{2}\to1}}\frac{\partial^{2}}{\partial r_{2}\partial r_{1}}\frac{\tr\mathbf{X}(r_{1})\tr\mathbf{X}(r_{2})}{\Gamma(mn+r_{1}+r_{2})}, \\
T_{c} &=& -\Gamma(mn)\lim_{\substack{r_{1}\to1\\r_{2}\to1}}\frac{\partial^{2}}{\partial r_{2}\partial r_{1}}\frac{\tr\left(\mathbf{X}(r_{1})\mathbf{X}(r_{2})\right)}{\Gamma(mn+r_{1}+r_{2})}.
\end{eqnarray}

\subsection{Calculating $T_{a}$}
\begin{eqnarray}
T_{a}&=&\Gamma(mn)\lim_{\substack{r_{1}\to1\\r_{2}\to1}}\frac{\partial^{2}}{\partial r_{2}\partial r_{1}}\frac{\tr\mathbf{X}(r_{1}+r_{2})}{\Gamma(mn+r_{1}+r_{2})} \\
&=&\frac{\Gamma(mn)}{\Gamma(mn+2)}\left(c_{3}-2c_{2}\psi_{0}(mn+2)+c_{1}\psi_{0}^{2}(mn+2)-c_{1}\psi_{1}(mn+2)\right) \nonumber,
\end{eqnarray}
where
\begin{eqnarray}
c_{1} &=& \tr\mathbf{X}(2), \\
c_{2} &=& \tr^{\prime}\mathbf{X}(2), \\
c_{3} &=& \tr^{\prime\prime}\mathbf{X}(2).
\end{eqnarray}
Here, for example, we have used the shorthand notation
\begin{equation}
c_{2}=\frac{\partial}{\partial r_{1}}\tr\mathbf{X}(r_{1}+r_{2})\bigg|_{\substack{r_{1}=1\\r_{2}=1}}:=\tr^{\prime}\mathbf{X}(2).
\end{equation}
Computing the coefficients $c_{1}$, $c_{2}$, and $c_{3}$ consists of taking up to the second derivatives of~(\ref{eq:xij}), resolving the indeterminacy of limits by using series expansion of gamma and polygamma functions around negative integers
\begin{eqnarray}
\Gamma(-l+\epsilon)&=&\frac{(-1)^{l}}{l!\epsilon}\left(1+\psi_{0}(l+1)\epsilon+o\left(\epsilon^2\right)\right),\\
\psi_{0}(-l+\epsilon)&=&-\frac{1}{\epsilon}\left(1-\psi_{0}(l+1)\epsilon+o\left(\epsilon^2\right)\right),\\
\psi_{1}(-l+\epsilon)&=&\frac{1}{\epsilon^2}\left(1+o\left(\epsilon^2\right)\right),
\end{eqnarray}
and evaluating the summations of resulting polygamma functions. These types of computation have been performed in~\cite{Wei17}, which similarly lead to the results
\begin{eqnarray}
c_{1} &\!\!\!\!\!\!=\!\!\!\!\!\!& mn(m+n), \\
c_{2} &\!\!\!\!\!\!=\!\!\!\!\!\!& \frac{m}{6}\left(6n(m+n)\psi_{0}(n+1)+m(m+9n-3)-3n+2\right), \\
c_{3} &\!\!\!\!\!\!=\!\!\!\!\!\!& 2mn(m+n)\sum_{k=1}^{m}\frac{\psi_{0}(k+n-m)}{k}+\nonumber \\
&&\!\!\!\!\!\!\!\!\!\!\!\!\frac{n}{3}\left(9mn-3m+n^2-3n+2\right)\psi_{0}(n)+2mn(m+n)\psi_{0}(1)\psi_{0}(n-m)+\nonumber \\
&&\!\!\!\!\!\!\!\!\!\!\!\!\frac{1}{3}\left(m^3+9m^{2}n+3m^{2}-9mn^{2}+2m-n^{3}-3n^{2}-2n\right)\psi_{0}(n-m)-\nonumber \\
&&\!\!\!\!\!\!\!\!\!\!\!\!mn(m+n)\left(\psi_{0}(n-m)^{2}+2\psi_{0}(n-m)\left(\psi_{0}(m)-\psi_{0}(n)\right)-\psi_{1}(n-m)\right)-\nonumber \\
&&\!\!\!\!\!\!\!\!\!\!\!\!\frac{m}{18}\left(5 m^2+75 m n+3 m+6 n^2+3 n-8\right).\label{eq:c3}
\end{eqnarray}
Note that in obtaining the above results one has to make use of sum identities of powers of digamma functions, e.g.,~\cite[Eqs.~(A4)-(A6)]{Wei17} and a semi closed-form sum identity~\cite[Eq.~(A12)]{Wei17}. Mathematica is currently unable to evaluate these two types of summations~\cite{KG}.

\subsection{Calculating $T_{b}$}
\begin{eqnarray}
T_{b}&=&\Gamma(mn)\lim_{\substack{r_{1}\to1\\r_{2}\to1}}\frac{\partial^{2}}{\partial r_{2}\partial r_{1}}\frac{\tr\mathbf{X}(r_{1})\tr\mathbf{X}(r_{2})}{\Gamma(mn+r_{1}+r_{2})} \\
&=&\frac{\Gamma(mn)}{\Gamma(mn+2)}\left(c_{5}^{2}-2c_{4}c_{5}\psi_{0}(mn+2)+c_{4}^{2}\psi_{0}^{2}(mn+2)-c_{4}^{2}\psi_{1}(mn+2)\right) \nonumber,
\end{eqnarray}
where
\begin{eqnarray}
c_{4} &=& \tr\mathbf{X}(1), \\
c_{5} &=& \tr^{\prime}\mathbf{X}(1).
\end{eqnarray}
Similarly as for $T_{a}$, the coefficients $c_{4}$ and $c_{5}$ are computed as
\begin{eqnarray}
c_{4} &=& mn, \\
c_{5} &=& mn\psi_{0}(n)+\frac{m}{2}(m+1).
\end{eqnarray}

\subsection{Calculating $T_{c}$}
\begin{eqnarray}
T_{c}&=&-\Gamma(mn)\lim_{\substack{r_{1}\to1\\r_{2}\to1}}\frac{\partial^{2}}{\partial r_{2}\partial r_{1}}\frac{\tr\left(\mathbf{X}(r_{1})\mathbf{X}(r_{2})\right)}{\Gamma(mn+r_{1}+r_{2})} \\
&=&\frac{\Gamma(mn)}{\Gamma(mn+2)}\left(c_{6}\psi_{1}(mn+2)-c_{6}\psi_{0}^{2}(mn+2)+2c_{7}\psi_{0}(mn+2)-c_{8}\right) \nonumber,
\end{eqnarray}
where the coefficients
\begin{eqnarray}
c_{6} &=& \tr\mathbf{X}^{2}(1), \\
c_{7} &=& \tr\mathbf{X}^{\prime}(1)\mathbf{X}(1), \\
c_{8} &=& \tr\left(\mathbf{X}^{\prime}(1)\right)^{2},
\end{eqnarray}
are similarly computed as
\begin{eqnarray}
c_{6} &\!\!\!\!\!\!=\!\!\!\!\!\!& mn(m+n-1), \\
c_{7} &\!\!\!\!\!\!=\!\!\!\!\!\!& mn(m+n-1)\psi_{0}(n+1)+\frac{m}{6}(m-1)(m+9n-5), \\
c_{8} &\!\!\!\!\!\!=\!\!\!\!\!\!& 2mn(m+n)\sum_{k=1}^{m}\frac{\psi_{0}(k+n-m)}{k}+\nonumber \\
&&\!\!\!\!\!\!\!\!\!\!\!\! \frac{1}{3}\left(-3m^2+9mn^2-9mn-3m+n^3-3n^2+2n\right)\psi_{0}(n)-mn\psi_{0}^{2}(n)+\nonumber \\
&&\!\!\!\!\!\!\!\!\!\!\!\! \frac{1}{3}\left(m^3+9m^{2}n+3m^{2}-9mn^{2}+2m-n^{3}-3n^{2}-2n\right)\psi_{0}(n-m)+\nonumber \\
&&\!\!\!\!\!\!\!\!\!\!\!\! mn(m+n)\big(\!-\psi_{0}^{2}(n-m)+2(\psi_{0}(n)-\psi_{0}(m)+\psi_{0}(1))\psi_{0}(n-m)+\nonumber \\
&&\!\!\!\!\!\!\!\!\!\!\!\! \psi_{1}(n-m)-\psi_{1}(n)\big)-\frac{m}{18}\left(5m^2+75mn+12m+6n^2+3n+1\right).\label{eq:c8}
\end{eqnarray}

\subsection{Recovering $\mathbb{V}\!_{f}\!\left[S\right]$}
Putting the results together, we arrive at the claimed formula~(\ref{eq:V}),
\begin{eqnarray}
\mathbb{V}\!_{f}\!\left[S\right] &=& \mathbb{E}_{f}\!\left[S^{2}\right]-\mathbb{E}_{f}^{2}\!\left[S\right] \\
&=& \frac{\Gamma(mn)}{\Gamma(mn+2)}\big(c_{5}^{2}-c_{8}+c_{3}+2\left(c_{7}-c_{4}c_{5}-c_{2}\right)\psi_{0}(mn+2)+\nonumber \\
&&\left(c_{4}^{2}-c_{6}+c_{1}\right)\left(\psi_{0}^{2}(mn+2)-\psi_{1}(mn+2)\right)\!\big)-\nonumber \\
&&\left(\psi_{0}(mn+1)-\psi_{0}(n)-\frac{m+1}{2n}\right)^{2} \label{eq:c38} \\
&=&-\psi_{1}\left(mn+1\right)+\frac{m+n}{mn+1}\psi_{1}\left(n\right)-\frac{(m+1)(m+2n+1)}{4n^{2}(mn+1)}
\end{eqnarray}
after some necessary simplification by the identities
\begin{eqnarray}
\psi_{0}(l+n)&=&\psi_{0}(l)+\sum_{k=0}^{n-1}\frac{1}{l+k}, \\
\psi_{1}(l+n)&=&\psi_{1}(l)-\sum_{k=0}^{n-1}\frac{1}{(l+k)^2}.
\end{eqnarray}

The essential derivation of the variance resides in the tedious calculations of the coefficients $c_{1}$ to $c_{8}$, which were computed in a different order in the proof~\cite{Wei17}. In particular, one of the key ingredients in the proof was to capture the cancellations of an unsimplifiable term
\begin{equation}
\sum_{k=1}^{m}\frac{\psi_{0}(k+n-m)}{k},
\end{equation}
which is also observed in the present note, cf.~(\ref{eq:c3}),~(\ref{eq:c8}), and~(\ref{eq:c38}). Capturing the cancellations requires certain tailor-made simplification tools derived in~\cite{Wei17}, which are unavailable in the computer algebra system Mathematica~\cite{KG}. It is therefore unclear the statement in~\cite{Bianchi19} that the calculations of~(\ref{eq:T2vN}) were performed with the help of Mathematica to yield the final result~(\ref{eq:V}).

\end{document}